\documentstyle[11pt,epsfig]{article}
\title{\bf Effect of the Chameleon Scalar Field on Brane Cosmological Evolution}
\author{ Y. Bisabr\footnote{email:
y-bisabr@srttu.edu}~~ and~~ F. Ahmadi\footnote{email:fahmadi@srttu.edu} \\{\small Department of Physics, Shahid Rajaee
Teacher Training University, Lavizan, Tehran 16788, Iran.}\\}
\begin{document}
\maketitle
\begin{abstract}
We have investigated a brane world model in which the gravitational
field in the bulk is described both by a metric tensor and a
minimally coupled scalar field.  This scalar field is taken to be a
chameleon with an appropriate potential function.  The scalar field
interacts with matter and there is an energy transfer between the
two components. We find a late-time asymptotic solution which
exhibits late-time accelerating expansion. We also show that the
Universe recently crosses the phantom barrier without recourse to
any exotic matter. We provide some thermodynamic arguments which
constrain both the direction of energy transfer and dynamics of the
extra dimension.
\end{abstract}
\vspace{.7cm}
\section{Introduction}
General Relativity has brilliant successes in explaining
gravitational phenomena in Solar System.  It is also a powerful tool
to explain theoretically many observational facts about the Universe
such as expansion of the universe, light element abundances and
gravitational waves. Despite all the successes, there are also some
unresolved problems such as inflation, the cosmological constant
problem and the problems associated with the dark sector, i.e., dark
matter and dark energy. These problems have motivated people to seek
for some modifications of the theory. Among many possibilities,
there are models that deal with extra dimensions. Most of these
models propose that our four-dimensional world is a hypersurface (or
brane) embedded in a higher dimensional space-time (or bulk). The
gravitational field propagates into the bulk while matter systems or
standard fields are confined to live in the brane. The most
well-known model in this context is the model proposed by Randall
and Sundrum (RS). In the so-called RSI model \cite{ran}, they
proposed a mechanism to solve the hierarchy problem with use of two
branes, while in the RSII model \cite{sun} they considered a single
brane with a positive tension. In the latter model, the
extra-dimension is compactified and a four-dimensional Newtonian
gravity is recovered at low energies. The cosmological evolution of
such a brane world scenario has been extensively investigated and
modifications of the gravitational equations have been studied
\cite{brax, bin}.\\
The basic idea in the brane world models can be extended to
scalar-tensor brane models in which gravity in the bulk is described
by a five-dimensional spacetime metric together with a scalar field
( see for instance \cite{Maeda} ). There are different motivations
for introducing a bulk scalar field in brane world scenarios.  This
scalar field may be used to formulate a low-energy effective theory
\cite{sc} or to address the gauge hierarchy problem \cite{yang}. One
of the important motivations to introduce such a bulk scalar field
is to stabilize the distance between the two branes in the RSI model
\cite{Kanti}.  Another explored possibility is formulating inflation
without an inflaton field on the brane \cite{Hime}. It is also shown
that decaying of the bulk
scalar field can lead to entropy production \cite{Yoko}.\\
 There is also a recent tendency in the Literature
\cite{kh} to interpret the scalar field as a chameleon field
\cite{cham}.  In these chameleon brane world models the scalar field
interacts with the matter system via the metric tensor and it is
assumed that it can be heavy enough in the environment of the
laboratory tests so that the local gravity constraints are
satisfied.  Meanwhile, it can be light enough in the low-density
environment to be considered as cosmologically viable. In the
present work we will investigate such a gravitational model with the
assumption that the scalar field has a minimal coupling with gravity
in the bulk.  We will focus on the late-time behavior of the
Universe and show that even though the scalar field is normal in the
sense that its energy-momentum tensor satisfies weak energy
condition, it causes the Universe to cross the phantom boundary.\\
The present work is organized as follows: In section 2, we introduce
the chameleon brane world model and derive the field equations.  In
section 3, we write the field equations for a five-dimensional
metric and then induce them on the brane with appropriate boundary
conditions.  In section 4, we will consider some cosmological
aspects of the model.  We first show that due to interaction of the
scalar field and matter evolution of both corresponding energy
densities are modified.  One immediate implication of such a
modification is an energy transfer between the two components. In
our analysis the chameleon actually appears as a normal field
satisfying weak energy condition. By finding a late-time asymptotic
solution for the field equations we will show that the Universe
suffers a late-time accelerating expansion and a recent cross-over
from a decelerated to an accelerated phase. In section 5, we provide
some thermodynamic arguments for the interaction process.  In
section 6, we draw our conclusions.
~~~~~~~~~~~~~~~~~~~~~~~~~~~~~~~~~~~~~~~~~~~~~~~~~~~~~~~~~~~~~~~~~~~~~~~~~~~~~~~~~~~~~~~~~~~~~~~~~~~~~~~~~~~~~~~~~~~~~~~~~~~~~~~~~~~~~~~~~~~~~~~
\section{The Model}
 We consider the following action\footnote{We work in the unit system in which $k_5=1$.}
\begin{equation}
S=\frac{1}{2}\int d^{5}x \sqrt{-g}\left[R - g^{AB}\nabla_{A}\phi
\nabla_{B}\phi -2 V(\phi)\right]+\int d^{4}x L_{m}(\psi_{m}
 \bar{h}_{\mu\nu})
\label{eq1}
\end{equation}
where the first term is the five-dimensional gravity in the presence
of a minimally coupled scalar field $\phi$. The second term is the
action of some matter fields on the brane which is taken to be
coupled to the scalar field via
$\bar{h}_{\mu\nu}=A^{2}(\phi)h_{\mu\nu}$ with $h_{\mu\nu}$ and
$\bar{h}_{\mu\nu}$\footnote{Latin indices denote 5-dimensional
components $A, B, ...=0, .., 4$ while Greek indices run over
four-dimensional brane $\mu, \nu,... = 0, ..., 3$ and $y$ is the
coordinate transverse to the brane.} being four-dimensional metrics
on the brane.
\\Varying the action with respect to the metric $g^{AB}$, gives
\begin{equation}
G_{AB}=\left[T_{AB}|_{bulk}+T_{AB}|_{brane}\right] \label{eq2}
\end{equation}
where
\begin{equation}
T_{AB}|_{bulk}=\nabla_{A}\phi\nabla_{B}\phi-\frac{1}{2}g_{AB}\nabla_{C}\phi\nabla^{C}\phi-g_{AB}V(\phi)
\label{eq3}
\end{equation}
and
\begin{equation}
T_{AB}|_{brane}=\delta^{\mu}_{A}\delta^{\nu}_{B}
\tau_{\mu\nu}\frac{\delta(y)}{b}
\label{eq4}
\end{equation}
Here we take $g_{AB}dz^{A}dz^{B}=h_{\mu\nu}dx^{\mu}dx^{\nu}+b^{2}(t,y)dy^{2}$
and $\tau_{\mu\nu}=A^{2}(\phi)\tau^{m}_{\mu\nu}$ with
$\tau^{m}_{\mu\nu}=\frac{-2}{\sqrt{-\bar{h}}}\frac{\delta
L_{m}}{\delta \bar{h}^{\mu\nu}}$. We will consider $\tau_{\mu\nu}$
as the stress-tensor of a perfect fluid with energy density
$\rho_{b}$ and pressure $P_{b}$. Variation of the action with
respect to $\phi$, leads to
\begin{equation}
\Box\phi-\frac{dV}{d\phi}=-\beta(\phi)T|_{brane} \label{eq5}
\end{equation}
where $\beta(\phi)=\frac{d \ln A(\phi)}{d \phi}$. By applying
Bianchi identities to (\ref{eq2}), we obtain
\begin{equation}
\nabla_{A}T^{AB}|_{brane}=-\nabla_{A}T^{AB}|_{bulk}=\beta(\phi)T|_{brane}\nabla^{B}\phi
\label{eq6}
\end{equation}
~~~~~~~~~~~~~~~~~~~~~~~~~~~~~~~~~~~~~~~~~~~~~~~~~~~~~~~~~~~~~~~~~~~~~~~~~~~~~~~~~~~~~~~~~~~~~~~~~~~~~~~~~~~~~~~~~~~~~~~~~~~~~~~~~~
\section{The brane-world paradigm}
We use the five-dimensional metric
\begin{eqnarray}
dS^{2}&=&h_{\mu\nu}dx^{\mu}dx^{\nu}+b^{2}(t,y)dy^{2}\nonumber\\&=&
-\tilde{n}^{2}(t,y)dt^{2}+\tilde{a}^{2}(t,y)[\frac{dr^{2}}{(1-kr^{2})}+r^{2}(d\theta^{2}+\sin^{2}\theta
d\phi^{2})]+\tilde{b}^{2}(t,y)dy^{2} \label{eq7}
\end{eqnarray}
with $k=0, +1, -1$. The metric coefficients are subjected to the
conditions
\begin{equation}
\tilde{n}(t,y)|_{brane}=1,\hspace{.75cm}\tilde{a}(t,y)|_{brane}=a(t),\hspace{.75cm}\tilde{b}(t,y)|_{brane}=b(t)
\label{eq8}
\end{equation}
with $a(t)$ being the scale factor. To write the bulk field
equations in compact form, we define \cite{bin}
\begin{equation}
F(t,y)\equiv\frac{(\tilde{a}'\tilde{a})^{2}}{\tilde{b}^{2}}-\frac{(\dot{\tilde{a}}\tilde{a})^{2}}{\tilde{n}^{2}}-k\tilde{a}^{2}
\label{eq9}
\end{equation}
where a prime denotes a derivative with respect to $y$. The (0,0)
and (5,5) components of the field equations become
\begin{equation}
F'=\frac{2\tilde{a}'\tilde{a}^{3}}{3}T^{0}_{0}|_{bulk} \label{eq10}
\end{equation}
\begin{equation}
\dot{F}=\frac{2\dot{\tilde{a}}\tilde{a}^{3}}{3}T^{5}_{5}|_{bulk}
\label{eq11}
\end{equation}
If we take
\begin{equation}
T^{A}_{B}|_{bulk}=diag[-\rho_{\phi}, P_{\phi}, P_{\phi}, P_{\phi},
P_{T}]
 \label{eq12}
\end{equation}
and assume that $\phi$ and therefore
$T^{0}_{0}|_{bulk}=-\rho_{\phi}$ are independent of $y$,  then we
can integrate (\ref{eq10}) which gives
\begin{eqnarray}
F-\frac{1}{6}\tilde{a}^{4}T^{0}_{0}|_{bulk}+C_1=F+\frac{1}{6}\tilde{a}^{4}\rho_{\phi}+C_1=0
\label{eq13}
\end{eqnarray}
where $C_1$ is a constant of integration. Since $\phi$ is only
time-dependent, we have
\begin{equation}
T^{0}_{0}|_{bulk}=\frac{1}{2}\nabla_{0}\phi\nabla^{0}\phi-V(\phi)
\label{eq14}
\end{equation}
\begin{equation}
T^{5}_{5}|_{bulk}=-\frac{1}{2}\nabla_{0}\phi\nabla^{0}\phi-V(\phi)
\label{eq15}
\end{equation}
This results in
$T^{0}_{0}|_{bulk}-T^{5}_{5}|_{bulk}=\nabla_{0}\phi\nabla^{0}\phi$.
From time-derivative of (\ref{eq10}) and derivative of (\ref{eq11})
with respect to $y$, one then finds
\begin{equation}
\frac{d}{dt}T^{0}_{0}|_{bulk}=-\frac{(\frac{d\tilde{a}^{4}}{dt})(\nabla_{0}\phi\nabla^{0}\phi)}{\tilde{a}^{4}}
\label{eq16}
\end{equation}
Using this equation and (\ref{eq13}), we arrive at
\begin{equation}
\dot{F}=\frac{2}{3}\dot{\tilde{a}}\tilde{a}^{3}T^{5}_{5}|_{bulk}-\frac{dC_1}{dt}
\label{eq17}
\end{equation}
Comparing this with (\ref{eq11}), indicates that $C_1$ is
time-independent. From (\ref{eq9})
 and (\ref{eq13}), we can write
\begin{equation}
(\frac{\dot{\tilde{a}}}{\tilde{n}\tilde{a}})^{2}=\frac{1}{6}\rho_{\phi}+(\frac{\tilde{a}'}{\tilde{b}\tilde{a}})^{2}
-\frac{K}{\tilde{a}^{2}}+\frac{C_1}{\tilde{a}^{4}} \label{eq18}
\end{equation}
For inducing the field equations on the brane, one usually uses the
junction conditions. They simply relate the jumps of derivative of
the metric across the brane to the stress-energy tensor inside the
brane. To do this, we first note that homogeneity and isotropy imply
that
\begin{equation}
T^{A}_{B}|_{brane}=\frac{\delta(y)}{b}diag[-\rho_{b}, P_{b}, P_{b}, P_{b}, 0]
\label{eq19}
\end{equation}
and
\begin{equation}
T^{\mu}_{\nu}|_{brane}(x^{\alpha},0)=\lim_{\epsilon\rightarrow 0}
\int^{\frac{\epsilon}{2}}_{-\frac{\epsilon}{2}}T^{\mu}_{\nu}|_{brane}
b dy=\tau^{\mu}_{\nu}(x^{\alpha})
 \label{eq20}
\end{equation}
where the energy density $\rho_{b}$ and pressure $P_{b}$ are only
functions of time. The
 junction condition is
\begin{equation}
\frac{[\tilde{a}']}{a b}=-\frac{1}{3}\rho_{b}\hspace{.5cm}
\label{eq21}
\end{equation}
where $[Q]=Q(0^{+})-Q(0^{-})$ denotes the jump of function $Q$
across $y=0$. Assuming the symmetry $y\leftrightarrow -y$, the
 generalized Friedmann equation becomes
\begin{equation}
H^{2}=\frac{1}{6}\rho_{\phi}+\frac{1}{36}\rho_{b}^{2}+\frac{C_1}{a^{4}}-\frac{k}{a^{2}}
\label{eq22}
\end{equation}
where $H\equiv\frac{\dot{a}}{a}$ is the Hubble parameter.
~~~~~~~~~~~~~~~~~~~~~~~~~~~~~~~~~~~~~~~~~~~~~~~~~~~~~~~~~~~~~~~~~~~~~~~~~~~~~~~~~~~~~~~~~~~~~~~~~~~~~~~~~~~~~~~~~~~~~~~~~~~~~~~~~~~
\section{Cosmological Implications}
In this section we will consider some cosmological implications of
the above model.
\subsection{The conservation equation on the brane} We would like to consider a class of solutions of the field
equations (\ref{eq2}) under the assumption that the metric
coefficients in (\ref{eq7}) are separable functions of their
arguments \cite{pon}. In this class, we have
\begin{equation}
\tilde{n}(t,y)=n(y), \hspace{.5cm}
\tilde{a}(t,y)=a(t)Y(y),\hspace{.5cm} \tilde{b}(t,y)=b(t)
\label{eq23}
\end{equation}
together with $Y(y)|_{brane}=Y(0)=1$ and $n(y)|_{brane}=n(0)=1$.
From $G_{05}=0$, it follows that
\begin{equation}
(\frac{n'}{n})=(1-s)(\frac{Y'}{Y}), \hspace{1cm}
\frac{\dot{b}}{b}=s\frac{\dot{a}}{a} \label{eq24}
\end{equation}
where $s$ is an arbitrary constant. This leads to a relation between
$a(t)$ and $b(t)$, namely $b(t)=C_2 a^{s}$ with $C_2$ being a constant of
integration.\\
There is a constraint on the parameter $s$  coming from arguments
related to temporal variation of the gravitational coupling. These
arguments lead to $(\frac{\dot{G}}{G})=-sH$ \cite{ran}
\cite{sun}\footnote{If spacetime has one spatial extra dimension,
then there will be a relation such as $b G= G_* $ \cite{ar} where
$G$ and $G_*$ are, respectively, four and five dimensional
gravitational couplings and $b$ is radius of the extra dimension.
Then $\frac{\dot{G}}{G}=-\frac{\dot{b}}{b}=-s H$ where $G_*$ is
assumed to be a constant.}. On the other hand, observations on the
time variation of $G$ give $\frac{\dot{G}}{G}=gH$, with $g$ being
bounded by $\mid g
\mid \leq 0.1$ \cite{Uzan}. Thus the absolute value of $s$ is constrained to be $|s|\leq 0.1 $.\\
One can use the equation (\ref{eq20}) to write
 (\ref{eq6}) on the brane ($y=0$)
\begin{equation}
\dot{\rho}_{\phi}+3H(\omega_{\phi}+1)\rho_{\phi}+\frac{\dot{b}}{b}\dot{\phi}^2=Q
\label{24-1}\end{equation}
 \begin{equation}
\dot{\rho_{b}}+3H(\omega_{b}+1+\frac{s}{3})\rho_b=-Q \label{eq25}
\end{equation}
where $Q= \beta(\phi)(3\omega_{b}-1)\dot{\phi}\rho_b$. The solution
of the latter is
\begin{equation}
\rho_{b}=\rho_{0b}a^{-3(\omega_{b}+1+\frac{s}{3})}e^{(1-3\omega_{b})\int\beta
d\phi} \label{eq26}
\end{equation}
with $\rho_{0b}$ being an integration constant. This solution
indicates that the evolution of the matter density is modified due
to
 interaction with $\phi$. This expression can
 be also written as \cite{bis}
\begin{equation}
\rho_{b}=\rho_{0b}a^{-3(\omega_{b}+1+\frac{s}{3})+\epsilon}
\label{eq27}
\end{equation}
with $\epsilon$ being defined by
\begin{equation}
\epsilon=\frac{(1-3\omega_{b})\int \beta d\phi}{\ln a} \label{eq28}
\end{equation}
Before going further, we would like to show that contrary to the
usual dark energy fields the scalar field $\phi$ satisfies the weak
energy condition. To do this, we first use the relation (\ref{eq23})
to write (\ref{eq5}) on the brane
\begin{equation}
\ddot{\phi}+(3+s)H\dot{\phi}+\frac{dV}{d\phi}=\beta(\phi)(3\omega_{b}-1)\rho_{b}
\label{eq30}
\end{equation}
Moreover, equation (\ref{eq16}) on the brane gives
\begin{equation}
\ddot{\phi}+\frac{dV}{d\phi}=-4H\dot{\phi} \label{eq31}
\end{equation}
Combining these two equations leads to
\begin{equation}
(s-1)H\dot{\phi}=\beta(\phi)(3\omega_{b}-1)\rho_{b} \label{eq32}
\end{equation}
We then write (\ref{24-1}) in the following form
\begin{equation}
\dot{\rho_{\phi}}+3\frac{\dot{a}}{a}(\omega_{\phi}+1)\rho_{\phi}+\frac{\dot{a}}{a}\dot{\phi}^{2}=0,
\label{eq33}
\end{equation}
where (\ref{eq23}) and (\ref{eq32}) have been used.   From the
definition $\rho_{\phi}=\frac{1}{2}\dot{\phi}^{2}+V(\phi)$, we have
\begin{equation}
\dot{\rho}_{\phi}=(\ddot{\phi}+\frac{dV}{d\phi})\dot{\phi}
\label{eq34}
\end{equation}
Now combining (\ref{eq33}), (\ref{eq34}) and using (\ref{eq31})
gives
\begin{equation}
\dot{\phi}^{2}=(\omega_{\phi}+1)\rho_{\phi}
 \label{eq35}
\end{equation}
which means that $(\omega_{\phi}+1)>0$ and the scalar field $\phi$
satisfies the weak energy condition.
~~~~~~~~~~~~~~~~~~~~~~~~~~~~~~~~~~~~~~~~~~~~~~~~~~~~~~~~~~~~~~~~~~~~~~~~~~~~~~~~~~~~~~~~~~~~~~~~~~~~~~~~~~~~~~~~~~~~~~~~~~~~~~~~~~
\subsection{Late-time Behavior}
We are interested in late-time behavior of the Universe. To deal
with this issue we look for late-time asymptotic solutions of the
field equations.  When $t\rightarrow \infty$ (or
$a\rightarrow\infty$), equations (\ref{eq22}) and (\ref{eq30})
reduce to
\begin{equation}
H^{2}\approx
\frac{1}{6}\rho_{\phi}=\frac{1}{6}(\frac{1}{2}\dot{\phi}^2+V(\phi))
\label{aeq22}
\end{equation}
\begin{equation}
a^{-(s+3)} \frac{d}{dt}(\dot{\phi}a^{(s+3)})\approx
-\frac{dV(\phi)}{d\phi}\label{a-eq30}
\end{equation}
As a usual rule for solving this set of equations, one usually gives
the potential function $V(\phi)$ as an input and then finds the
functions $a(t)$ and $\phi(t)$.  However, we would like to follow a
different strategy. We will take $\dot{\phi}=a^n$, with $n$ being a
constant parameter, as an input and find $V(\phi)$ and $a(t)$ so
that the equations (\ref{aeq22}) and (\ref{a-eq30}) are satisfied.
The solutions are
\begin{equation}
a(t)=C^{-\frac{1}{n}}t^{-\frac{1}{n}}\label{aeq1}
\end{equation}
\begin{equation}
V(\phi)=V_0e^{-2C\phi}\label{aeq30}
\end{equation}
where $C=\frac{(-n)}{2\sqrt{3}}(\frac{-(s+3)}{n})^{\frac{1}{2}}$ and
$V_0=\frac{6(n+(s+3))}{n^2(s+3)}$. This set of solutions indicates
that the Universe is accelerating for $-1<n<0$. The functions $a(t)$
and $V(\phi)$ are plotted in Fig.1.
\begin{figure}[ht]
\begin{center}
\includegraphics[width=0.45\linewidth]{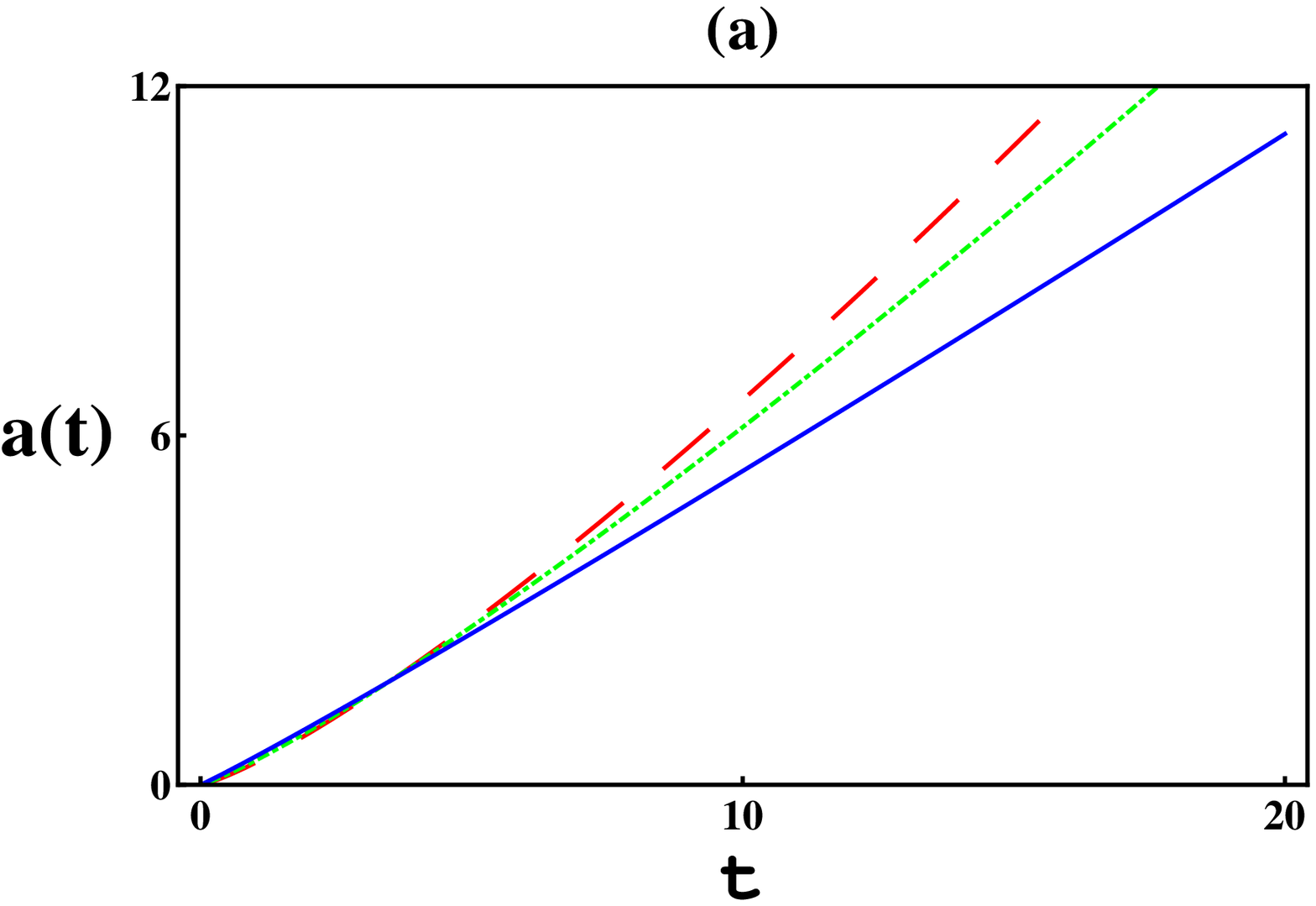}
\includegraphics[width=0.45\linewidth]{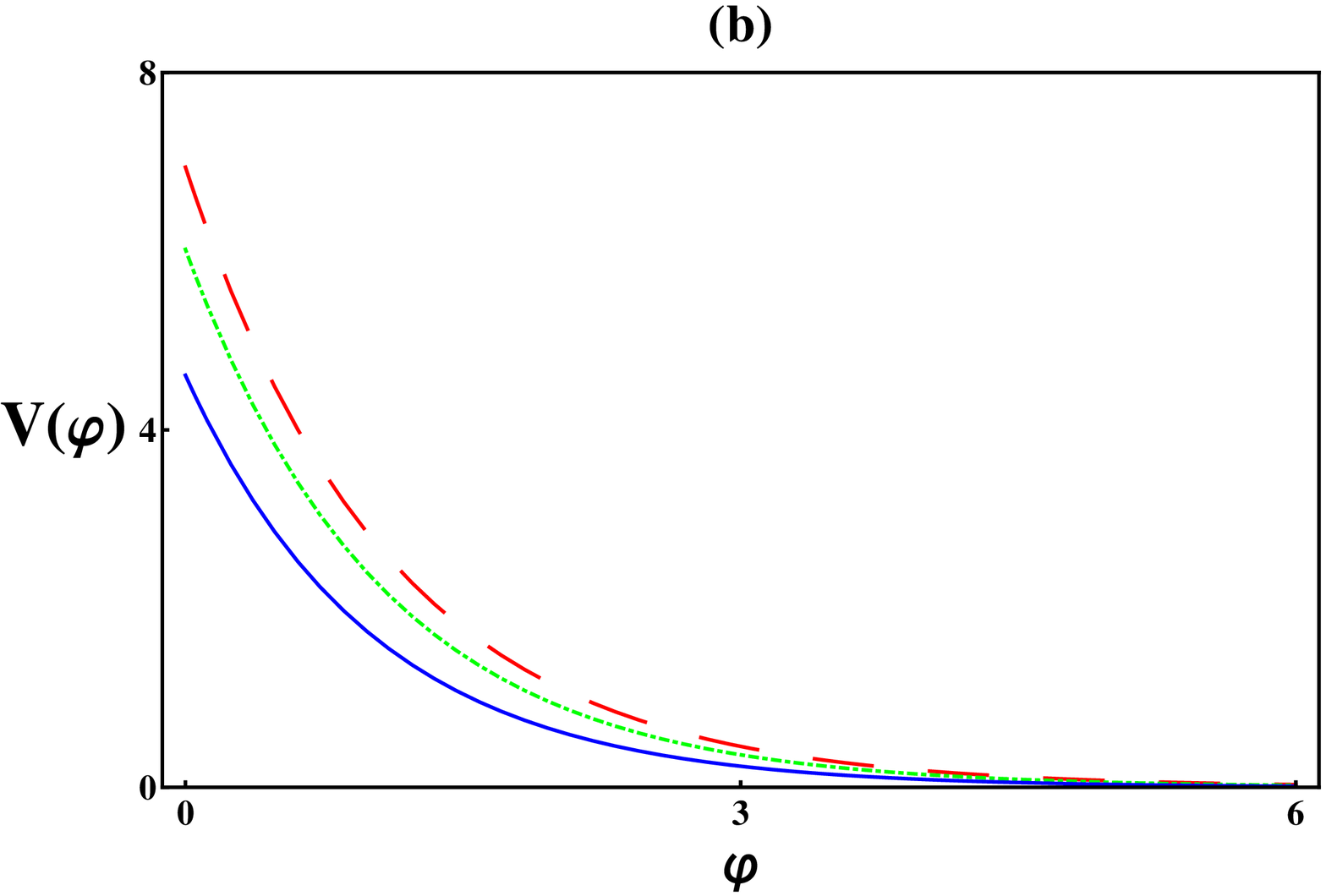}
\caption{The plot of $a(t)$ and $V(\phi)$ for  $n=-0.8$ and $s=0.08$
(dashed), $n=-0.85$ and $s=0.09$ (dotted) and $n=-0.95$ and $s=0.1$
(solid).}
\end{center}
\end{figure}
The figure shows that $V(\phi)$ has a run-away form as it should be
since $\phi$ is a chameleon field \cite{cham}.  \\The Universe has
not been in an accelerating phase at all the time and has suffered a
transition from an early decelerating phase to a recent accelerating
one.  To check that whether or nor the present model can generate
such a phase transition, we look at the effective equation of state
parameter $\omega_{eff}$.  We first re-write (\ref{eq25}) in the
form
\begin{equation}
\dot{\rho_{b}}+3H(\omega_{eff}+1)\rho_b=0 \label{aeq25}
\end{equation}
where
\begin{equation}
\omega_{eff}=\omega_b +\frac{s}{3}+\frac{Q}{3H\rho_b}
\end{equation}
$$
~~~~~~~~~~~~~~~~~~~~~~~~~~~~\omega_{eff}=\frac{s}{3}-\frac{1}{3}\beta(\phi)\frac{\dot{\phi}}{H}~~~~~~~~~~~~~~~~~~~~(\omega_b=0)
$$
Using the solution (\ref{aeq1}) and (\ref{aeq30}), gives
$\frac{\dot{\phi}}{H}=2\sqrt{3}(\frac{-n}{(s+3)})^{\frac{1}{2}}$
which is a constant.  This means that deceleration to acceleration
phase transition needs $\beta(\phi)$ not to be a constant.\\
Among many possible choices for $\beta(\phi)$, let us choose a
simple one $\beta(\phi)=\phi$ as an input coupling function. It
corresponds to $A(\phi)=e^{\frac{1}{2}\phi^2}$. The resulting
effective equation of state parameter is plotted in Fig.2.
\begin{figure}[ht]
\begin{center}
\includegraphics[width=0.45\linewidth]{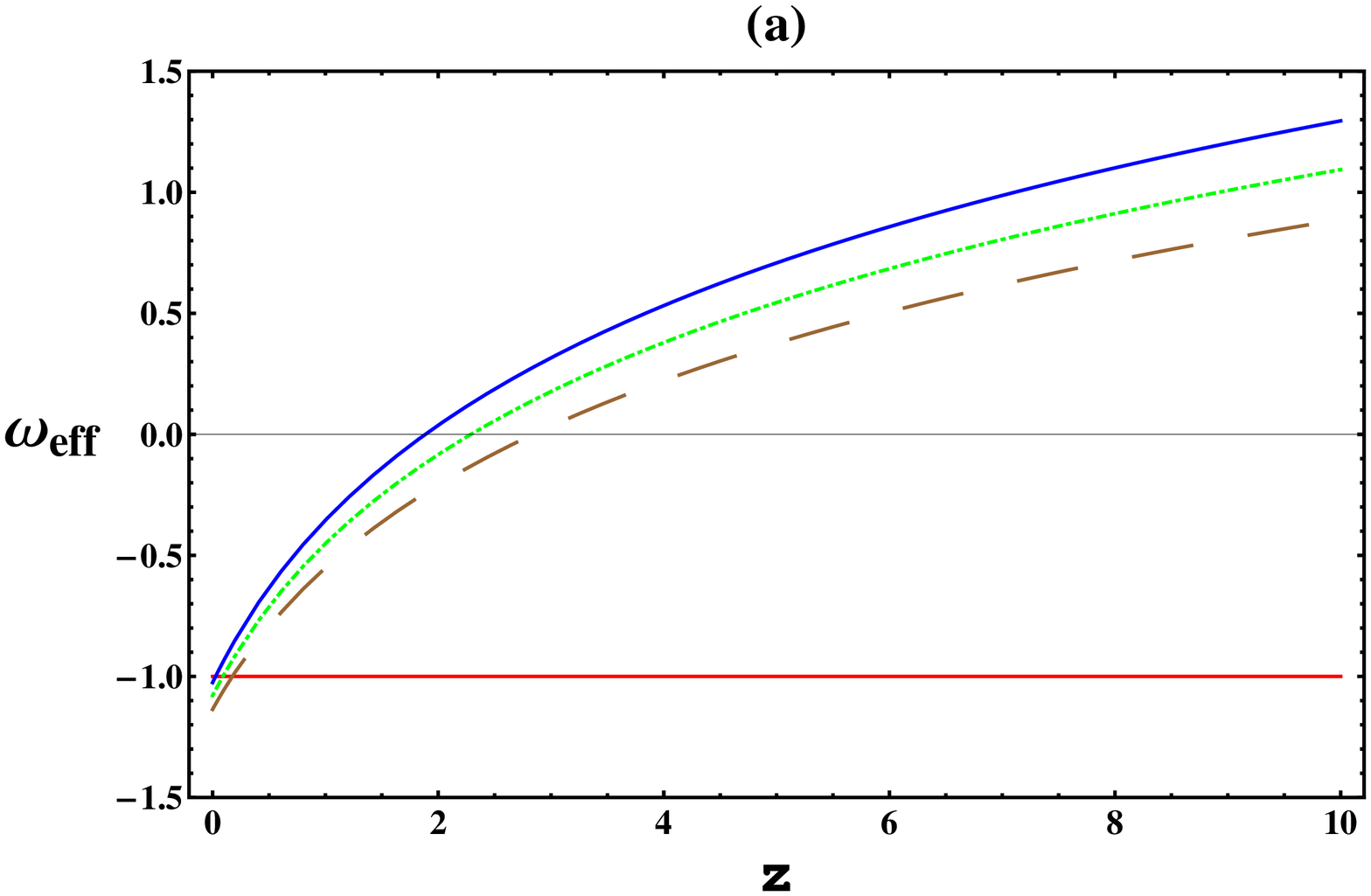}
\includegraphics[width=0.45\linewidth]{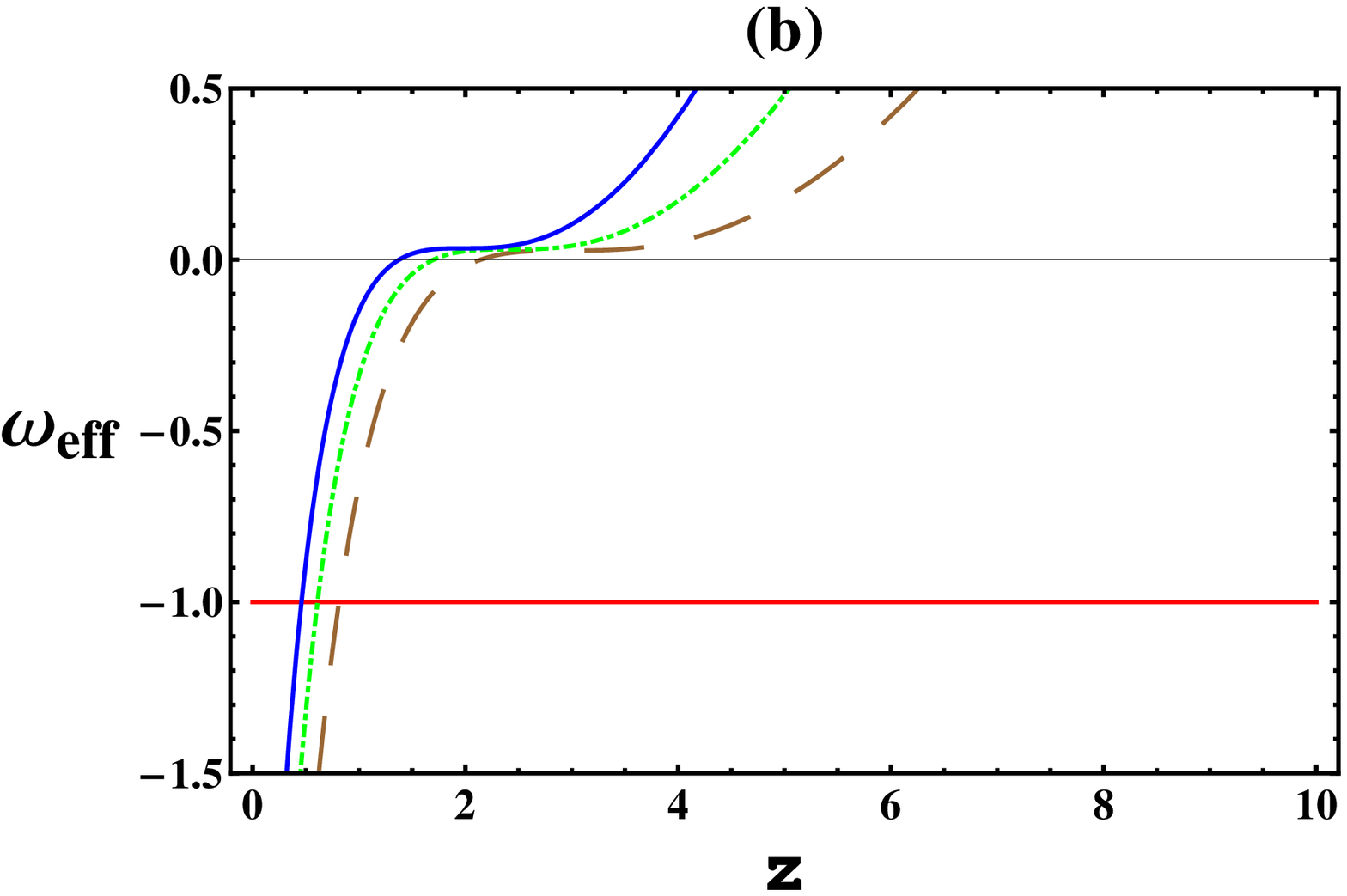}
\caption{The plot of $\omega_{eff}$ in terms of $z$ for
$\beta(\phi)=\phi$ panel (a) and $\beta(\phi)=\phi^3$ panel (b). The
curves correspond to $n=-0.65$ and $s=0.08$ (dashed), $n=-0.7$ and
$s=0.09$ (dotted) and $n=-0.75$ and $s=0.1$ (solid).}
\end{center}
\end{figure}
 As it is clear from the figure, the function
$\omega_{eff}$ exhibits a recent signature flip.  It also shows that
the Universe recently enters the phantom region.  \\The choice
$\beta(\phi)=\phi$ is not the only one that leads to a transition
from decelerating to accelerating phase.  The panel (b) of the Fig.2
shows $\omega_{eff}$ for another choice $\beta(\phi)=\phi^3 $. It
should be remarked that in both cases deceleration to acceleration
transition takes place when $\beta>0$ or $Q<0$. It means that in the
interacting process described by (\ref{24-1}) and (\ref{eq25}), the
direction of energy flow is so that matter is created. This seems to
be consistent with the results reported in \cite{p}.
~~~~~~~~~~~~~~~~~~~~~~~~~~~~~~~~~~~~~~~~~~~~~~~~~~~~~~~~~~~~~~~~
~~~~~~~~~~~~~~~~~~~~~~~~~~~~~~~~~~~~~~~~~~~~~
\section{Thermodynamic Analysis}
A thermodynamic description of a homogeneous and isotropic
interacting perfect fluid requires a knowledge of the particle flux
$N^{\alpha}=nu^{\alpha}$ and the entropy flux $S^{\alpha}=s
u^{\alpha}$ where  $n = N/a^3$, $s=n \sigma$ and $\sigma = S/N$ is
specific entropy (per particle) of the created or annihilated
particles. Since energy density of matter is given by $\rho_b=nM$
with $M$ being the mass of each particle, the appearance of the
extra term in the energy balance equation (\ref{eq25}) means that
this term can be attributed to a change of $n$ or $M$.  Here we
assume that the mass of each matter particle remains constant and
the extra term in the energy balance equation only leads to a change
of the number density $n$. In this case, the equation (\ref{eq25})
can be written as \footnote{Throughout this section we have set
$\omega_b=0$. }
\begin{equation}
\dot{n}+3H(1+\frac{s}{3})n=n\Gamma \label{c15}\end{equation} where
$\Gamma\equiv\beta(\phi)\dot{\phi}$ is the rate of creation (or
annihilation) of particles. The direction of energy transfer between
matter and the scalar field depends on the sign of $\Gamma$. If
$\Gamma>0$ (or $Q<0$), the energy goes inside of the matter system
and matter is created. If $\Gamma<0$ (or $Q>0$) the direction of
energy transfer is
reversed and matter is annihilated.\\
From $\sigma=S/N$, we have
\begin{equation}
\frac{\dot{\sigma}}{\sigma}=\frac{\dot{S}}{S}-\frac{\dot{N}}{N}\label{aba}
\end{equation}
With use of (\ref{c15}), the latter can be written as
\begin{equation}
\frac{\dot{S}}{S}=\frac{\dot{\sigma}}{\sigma}+(\Gamma-sH)
\label{aa}\end{equation} Since $n\propto a^{-3+(\epsilon-s)}$, the
total number of particles scale as $N\propto a^{\epsilon-s}$.  Thus
(\ref{aba})  can also be written as
\begin{equation}
\frac{\dot{S}}{S}=\frac{\dot{\sigma}}{\sigma}+(\epsilon-s)H
\label{bb}\end{equation} In an adiabatic process, when the overall
energy transfer is such that the specific entropy per particle
remains constant ($\dot{\sigma}=0$) \cite{li}, the second law of
thermodynamics ($\dot{S}\geq 0$) implies that $\epsilon-s\geq0$ in
an expanding Universe. In this case, when $\epsilon<0$ the parameter
$s$ is allowed to take only negative values. Alternatively speaking,
the extra dimension
shrinks with expansion of the Universe (see (\ref{eq24})).\\
In the non-adiabatic case, on the other hand, the second law of
thermodynamics requires that
\begin{equation}
\Gamma\geq
sH-\frac{\dot{\sigma}}{\sigma}
\end{equation}
which is also a constraint on the creation (or annihilation) rate
and evolution of the extra dimension.
~~~~~~~~~~~~~~~~~~~~~~~~~~~~~~~~~~~~~~~~~~~~~~~~~~~~~~~~~~~~~~~~~~~~~~~~~~~~~~~~~~~~~~~~~~~~~~~~~~~~~~~~~~~~~~~~~~~~~~~~~~~
\section{Conclusion}
We have investigated a brane world scenario in which gravity is
described by a five-dimensional metric together with a minimally
coupled scalar field.  The scalar field is a chameleon and interacts
with the matter sector. Due to this interaction the energy
associated with both the scalar field and matter system are not
separately conserved.  Thus evolution of matter energy density
modifies  and is controlled by $Q$.  When
$Q>0$ matter is created and energy is injecting into the matter
system so that the latter will dilute more slowly compared to
 its standard evolution $\rho_{b}\propto a^{-3(\omega_{b}+1)}$. On
the other hand, when $Q<0$ the reverse is true, namely that matter
is annihilated and the direction of energy transfer is outside of
the matter system (and into the scalar field) so that the
rate of dilution is faster than the standard one.\\
The main results of our analysis are the following:\\
1)  We have found a late-time asymptotic solution that exhibits
accelerating expansion. There is also a recent transition from a
decelerating phase to an accelerating one. \\
2) The interaction of chameleon field with matter plays an important
role in this phase transition. In order that this transition takes
place, the coupling function should be an evolving
function (or $\beta(\phi)$ should not be a constant). \\
3) Our analysis also indicates that the Universe has recently
entered the phantom region.  We emphasize that this behavior is not
attributed
to any exotic matter system.\\
4) A thermodynamic analysis  puts constraints  on $\Gamma$ and
evolution of the extra dimension in adiabatic and non-adiabatic
cases.
\\
There are some problems that are not investigated in the present
analysis such as behavior of the Universe at early times or the
cosmological constant problem.  They are deserved to be investigated elsewhere.

\end{document}